# New complete orthonormal sets of exponential type orbitals in standard convention and their origin

I.I. Guseinov

*Department of Physics, Faculty of Arts and Sciences, Onsekiz Mart University, Çanakkale, Turkey*

**Abstract**

In standard convention, the new complete orthonrmal sets of $\psi^{(\alpha*)}$-exponential type orbitals ($\psi^{(\alpha*)}$-ETOs) are introduced as functions of the complex or real spherical harmonics and $\mathcal{L}^{(\alpha*)}$-modified and $L^{(p*)}$-generalized Laguerre polynomials ($\mathcal{L}^{(\alpha*)}$-MPLs and $L^{(p*)}$-GLPs),

$$\psi^{(\alpha*)}_{nlm}(\zeta,\vec{r}) = (2\zeta)^{3/2} e^{-\frac{x}{2}} \mathcal{L}^{(\alpha*)}_{nl}(x) S_{lm}(\theta,\varphi)$$

$$\mathcal{L}^{(\alpha*)}_{nl}(x) = \left[\frac{(n-l-1)!}{(2n)^{\alpha*} \Gamma(q*+1)}\right]^{\frac{1}{2}} x^l L^{(p*)}_{n-l-1}(x),$$

where, $0 < \zeta < \infty$, $x = 2\zeta r$, $p* = 2l+2-\alpha*$, $q* = n+l+1-\alpha*$ and $\alpha*$ is the noninteger or integer (for $\alpha* = \alpha$) frictional quantum number ($-\infty < \alpha* < 3$). It is shown that the origin of the $\psi^{(\alpha*)}$-ETOs, $\mathcal{L}^{(\alpha*)}_{nl}$-MLPs and $L^{(p*)}$-GLPs is the self-frictional quantum forces which are analog of radiation damping or self-frictional forces introduced by Lorentz in classical electrodynamics. The relations for the quantum self-frictional potentials in terms of $\psi^{(\alpha*)}$-ETOs, $\mathcal{L}^{(\alpha*)}_{nl}$-MLPs and $L^{(p*)}$-GLPs, respectively, are established. We note that, in the case of disappearing frictional forces, the $\psi^{(\alpha*)}$-ETOs are reduces to the Schöringer's wave functions for the hydrogen-like atoms in standard convention and, therefore, become the noncomplete.

## 1. Introduction

In a previous paper [1], we have suggested in nonstandard convention the complete orthonormal sets of $\psi^{\alpha}$-ETOs with integer $\alpha$, where $-\infty < \alpha \leq 2$. It was shown that the Lambda and Coulomb-Sturmian ETOs introduced in Refs. [2-5] are the special cases of the $\psi^{\alpha}$-ETOs for $\alpha = 0$ and $\alpha = 1$, respectively, i.e., $\psi^0_{nlm} \equiv \Lambda_{nlm}$ and $\psi^1_{nlm} \equiv \psi_{nlm}$ (see also Refs. [6-9]). The purpose of this work is to construct the analytical relations for complete orthonormal sets of ETOs and quantum self-frictional potentials using generalized Laguerre polynomials defined in standard convention.

## 2. Definition and basic formulas

The complete orthonormal sets of $\psi^{(\alpha*)}$-ETOs with integer $(\alpha*=\alpha)$ and noninteger frictional quantum number $\alpha*$ in this work are defined as

$$\psi_{nlm}^{(\alpha*)}(\zeta,\vec{r}) = R_{nl}^{(\alpha*)}(\zeta,r) S_{lm}(\theta,\varphi) \tag{1}$$

$$R_{nl}^{(\alpha*)}(\zeta,r) = (2\zeta)^{\frac{3}{2}} R_{nl}^{(\alpha*)}(x) \tag{2}$$

$$R_{nl}^{(\alpha*)}(x) = e^{-\frac{x}{2}} \mathcal{L}_{nl}^{(\alpha*)}(x), \tag{3}$$

where

$$\mathcal{L}_{nl}^{(\alpha*)}(x) = \left[\frac{(n-l-1)!}{(2n)^{\alpha*}\Gamma(q*+1)}\right]^{\frac{1}{2}} x^l L_{n-l-1}^{(p*)}(x) \tag{4}$$

$$L_{n-l-1}^{(p*)}(x) = \frac{\Gamma(q*+1)}{(n-l-1)!\Gamma(p*+1)} {}_1F_1\left(-[n-l-1]; p*+1; x\right). \tag{5}$$

Here, $x=2\zeta r$, $p*=2l+2-\alpha*$, $q*=n+l+1-\alpha*$ and $-\infty<\alpha*<3$; $S_{lm}(\theta,\varphi)$ are the complex or real spherical harmonics; $L_{n-l-1}^{(p*)}$ and $\mathcal{L}_{nl}^{(\alpha*)}$ are the generalized and modified Laguerre polynomials, respectively. The ${}_1F_1(\eta;\gamma;x)$ is the confluent hypergeometric function determined as follows:

$$_1F_1(\eta;\gamma;x) = \sum_{k=0}^{\infty} \frac{(\eta)_k}{(\gamma)_k} \frac{x^k}{k!}, \tag{6}$$

where

$$(\eta)_k = \eta(\eta+1)\ldots(\eta+k-1) \text{ with } (\eta)_0 = 1 \tag{7}$$

is the Pochhammer symbol. In this work, the mathematical notation for $L^{(p*)}$-GLPs (see, for example, Ref. [11]) is used.

We notice that the Eqs. (1)-(3) are introduced by the use of method set out in previous works [1,10].

The orthonormality relations of $L^{(p*)}$-GLPs, $\mathcal{L}^{(\alpha*)}$-MLPs and $\psi^{(\alpha*)}$-ETOs are defined as follows:

$$\int_0^\infty e^{-x} x^{p*} L_{n-l-1}^{(p*)}(x) L_{n'-l-1}^{(p*)}(x) dx = \frac{\Gamma(q*+1)}{(n-l-1)!} \delta_{nn'} \tag{8}$$

$$\int_0^\infty e^{-x} x^2 \mathcal{L}_{nl}^{(\alpha*)}(x) \overline{\mathcal{L}}_{n'l}^{(\alpha*)}(x) dx = \delta_{nn'}, \qquad (9)$$

$$\int_0^\infty \psi_{nlm}^{(\alpha*)*}(\zeta,\vec{r}) \overline{\psi}_{n'l'm'}^{(\alpha*)}(\zeta,\vec{r}) d^3\vec{r} = \delta_{nn'}\delta_{ll'}\delta_{mm'}, \qquad (10)$$

where

$$\overline{\mathcal{L}}_{n'l}^{(\alpha*)}(x) = \left(\frac{2n}{x}\right)^{\alpha*} \mathcal{L}_{n'l}^{(\alpha*)}(x) \qquad (11)$$

$$\overline{\psi}_{nlm}^{(\alpha*)}(\zeta,\vec{r}) = \left(\frac{2n}{x}\right)^{\alpha*} \psi_{nlm}^{(\alpha*)}(\zeta,\vec{r}). \qquad (12)$$

The $L^{(p*)}$-GLPs satisfy the following differential equations [12]

$$\frac{d}{dx} L_{n-l-1}^{(p*)}(x) = -L_{n-l}^{(p*)}(x) \qquad (13)$$

$$x^2 \frac{d^2}{dx^2} L_{n-l-1}^{(p*)}(x) + (p*+1-x)\frac{d}{dx} L_{n-l-1}^{(p*)}(x) + (n-l-1) L_{n-l-1}^{(p*)}(x) = 0. \qquad (14)$$

By the use of Eqs. (3), (4), (13) and (14) it is easy to derive the following formulae:

for $\mathcal{L}_{nl}^{(\alpha*)}(x)$

$$\frac{d}{dx} \mathcal{L}_{nl}^{(\alpha*)}(x) = \frac{l}{x} \mathcal{L}_{nl}^{(\alpha*)}(x) - \frac{x}{\sqrt{[(n-l)q*]}} \mathcal{L}_{nl-1}^{(\alpha*)}(x) \qquad (15)$$

$$x \frac{d^2}{dx^2} \mathcal{L}_{nl}^{(\alpha*)}(x) + (p*+1-2l-x)\frac{d}{dx} \mathcal{L}_{nl}^{(\alpha*)}(x) + \left(n-1-\frac{l(p*-l)}{x}\right)\mathcal{L}_{nl}^{(\alpha*)}(x) = 0, \qquad (16)$$

for $R_{nl}^{(\alpha*)}(x)$

$$\frac{d}{dx} R_{nl}^{(\alpha*)}(x) = \left(-\frac{1}{2} + \frac{l}{x}\right) R_{nl}^{(\alpha*)}(x) - \frac{x}{\sqrt{[(n-l)q*]}} R_{nl-1}^{(\alpha*)}(x) \qquad (17)$$

$$x\frac{d^2}{dx^2} R_{nl}^{(\alpha*)}(x) + (3-\alpha*)\frac{d}{dx} R_{nl}^{(\alpha*)}(x) + \left(n+(1-\alpha*)\left(\frac{1}{2}-\frac{l}{x}\right) - \frac{l(l+1)}{x} - \frac{x}{4}\right) R_{nl}^{(\alpha*)}(x) = 0. \qquad (18)$$

In section 3, we need also to use the following relations:

$$\frac{dR_{nl}^{(\alpha*)}(x)}{dx} / R_{nl}^{(\alpha*)}(x) = -\frac{1}{2} + \frac{d\mathcal{L}_{nl}^{(\alpha*)}(x)}{dx} / \mathcal{L}_{nl}^{(\alpha*)}(x) \qquad (19)$$

$$\frac{dR_{nl}^{(\alpha*)}(x)}{dx}/R_{nl}^{(\alpha*)}(x) = -\frac{1}{2} + \frac{l}{x} + \frac{dL_{n-l-1}^{(p*)}(x)}{dx}/L_{n-l-1}^{(p*)}(x). \tag{20}$$

As we see that the definition of the $\psi^{(\alpha*)}$-ETOs and $\mathcal{L}_{nl}^{(\alpha*)}$-MLPs, which are the radial parts of $\psi^{(\alpha*)}$-ETOs, and derivation of their differential equations are based on the use of $L^{(p*)}$-GLPs.

## 3. Expressions for quantum frictional potentials

As we pointed out that the origin of the $\psi^{(\alpha*)}$-ETOs is the quantum frictional forces produced by the particles itself. Accordingly, the quantum self-frictional potential has to depend on the radial parts of $\psi^{(\alpha*)}$-ETOs the Schrödinger equation of which is defined by (see Ref.[10])

$$\left[-\frac{1}{x^2}\frac{d}{dx}\left(x^2\frac{d}{dx}\right) + \frac{l(l+1)}{x^2} + \frac{1}{4} + \frac{1}{2\zeta^2}V_{nl}^{(\alpha*)}(x)\right]R_{nl}^{(\alpha*)}(x) = 0. \tag{21}$$

This equation can be rewritten as the follows:

$$\left[x\frac{d^2}{dx^2} + (3-\alpha*)\frac{d}{dx} - \frac{(1-\alpha*)}{R_{nl}^{(\alpha*)}(x)}\frac{d}{dx} - \frac{l(l+1)}{x} - \frac{x}{4} - \frac{x}{2\zeta^2}V_{nl}^{(\alpha*)}(x)\right]R_{nl}^{(\alpha*)}(x) = 0. \tag{22}$$

Thus, we obtain for the $R_{nl}^{(\alpha*)}(x)$ two kinds of independent equations, Eqs. (18) and (22), one of which, Eq.(22), contains the quantum frictional potentials $V_{nl}^{(\alpha*)}(x)$. The comparison of Eqs. (18) and (22) gives

$$V_{nl}^{(\alpha*)}(x) = -\frac{2\zeta^2}{x}n - \frac{2\zeta^2}{x}(1-\alpha*)\left[\frac{dR_{nl}^{(\alpha*)}(x)}{dx}/R_{nl}^{(\alpha*)}(x) + \frac{1}{2} - \frac{l}{x}\right]. \tag{23}$$

Now we take into account Eqs. (17), (19) and (20). Then, we obtain for the quantum self-frictional potentials the following relations:

$$V_{nl}^{(\alpha*)}(\zeta,r) = U_n(\zeta,r) + U_{nl}^{(\alpha*)}(\zeta,r), \tag{24}$$

where

$$U_n(\zeta,r) = -\frac{\zeta n}{r} \tag{25}$$

$$U_{nl}^{(\alpha*)}(\zeta,r) = (1-\alpha*)\begin{cases} \left(2\zeta^2/\sqrt{(n-l)q*}\right)R_{nl-1}^{(\alpha*)}(x)/R_{nl}^{(\alpha*)}(x) & (26) \\ \left(2\zeta^2/\sqrt{(n-l)q*}\right)\mathcal{L}_{nl-1}^{(\alpha*)}(x)/\mathcal{L}_{nl}^{(\alpha*)}(x) & (27) \\ (\zeta/r)L_{n-l}^{(p*)}(x)/L_{n-l-1}^{(p*)}(x)\ . & (28) \end{cases}$$

Here, the function $U_n(\zeta,r)$ is the core attraction self-frictional potential. As we see that the total self-frictional potential $V_{nl}^{(\alpha*)}(\zeta,r)$ is a function of the self-frictional constant $\zeta$ and quantum numbers $n,l$ and $\alpha*$. The self-frictional parameter $\zeta$ can be chosen properly according to the nature of the particle and corresponding field under consideration.

It should be noted that the self-frictional properties disappear for $\zeta = \frac{Z}{n}$ and $\alpha* = \alpha = 1$, i.e.,

$$V_{nl}^{(\alpha*)}(\zeta,r) = -\frac{Z}{r} \qquad \text{for } \zeta = \frac{Z}{n} \text{ and } \alpha* = 1. \tag{29}$$

In this case, the $\psi^{(\alpha*)}$-ETOs are reduced to the Schrödinger's eigenfunction for the hydrogen-like atoms and become the noncomplete, i.e., $\psi_{nlm}^{(1)} \equiv \psi_{nlm}$, where $\psi_{nlm}$ is the Schrödinger's wave function in standard convention.

## 4. Conclusion

In this paper, by the use of explicit expression for the generalized Laguerre polynomials through the confluent hypergeometric series, the complete orthonormal sets of ETOs and self-frictional potentials are constructed. The relations for complete orthonormal sets of modified Laguerre polynomials through the GLPs are suggested. It is shown that the origin of the GLPs, therefore, of the ETOs and MLPs, is the self-frictional potentials of the field produced by the particle itself. The formulas for the $\psi^{(\alpha*)}$-ETOs, $\mathcal{L}_{nl}^{(\alpha*)}$-MLPs and $G^{(p*)}$-GLPs presented in this work can be useful tool in the numerous physical and mathematical applications. They can also be used for the wide applications in electronic structure calculations of atoms, molecules and solids.